\documentclass[twoside]{ilcws07}
\usepackage[dvips]{graphicx}
\usepackage{amssymb,amsmath,array}
\usepackage{units}
\usepackage{url}
\newcommand{\Lnc}{\ensuremath{\Lambda_{\text{NC}}} }
\newcommand{\ii}{\ensuremath{\textrm{i}} }
\newcommand{\ee}{\ensuremath{\textrm{e}} }

\begin{document}
\title{
The Noncommutative Standard Model at the ILC} 
\author{Ana Alboteanu\thanks{Speaker~\cite{url}}, Thorsten Ohl, and Reinhold R{\"u}ckl
\vspace{.3cm}\\
Universit{\"at} W{\"urzburg},
Institut f{\"ur} Theoretische Physik und Astrophysik \\
Am Hubland, 97074 W{\"urzburg}, Germany
}

\maketitle

\begin{abstract}
  We study phenomenological consequences of a noncommutative extension
  of the standard model in the $\theta$-expanded approach at
  the~ILC. We estimate the sensitivity of the~ILC for the
  noncommutative scale~$\Lnc$.  Comparing with earlier estimates for
  the~LHC, we demonstrate the complementarity of the experiments at
  the two colliders.
\end{abstract}

\section{The Model}
A noncommutative~(NC) structure of space-time
\begin{equation}\label{eq:NC-space-time}
  [\hat{x}^\mu,\hat{x}^\nu] = \ii \theta^{\mu\nu} = \ii\frac{C^{\mu\nu}}{\Lnc^2}\, 
\end{equation}
introduces a new energy scale $\Lnc$. The motivations
of~(\ref{eq:NC-space-time}) that are provided by string theory and
quantum gravity place this scale in the vicinity of the corresponding
Planck scale:~$\Lnc\approx M_{\textrm{Pl}}$.
If~$M_{\textrm{Pl}}\approx\unit[10^{19}]{GeV}$,
(\ref{eq:NC-space-time}) is unlikely to be ever probed directly by
collider experiments.  However, in models with additional space
dimensions~$M_{\textrm{Pl}}$ can be as low as the Terascale and, as a
result,~$\Lnc$ can be in the reach of future \unit{TeV} scale
colliders, like LHC and ILC.  Therefore, quantum field theories on NC
space-time~(NCQFT), in particular NC extensions of the standard model~(SM),
are interesting objects for collider phenomenology.  Using methods
developed for studying NCQFT at the
LHC~\cite{Alboteanu/Ohl/Rueckl:2006}, we have estimated the discovery
potential of the ILC and the sensitivity to the NC
parameters~(\ref{eq:NC-space-time}).

In this study, we assume a canonical structure of NC space-time,
i.\,e.~a constant antisymmetric $4\times 4$~matrix $C^{\mu\nu}$
in~(\ref{eq:NC-space-time}) that commutes with all the~$\hat x_\mu$.
For convenience, we parametrize~$C^{\mu\nu}$ in analogy to the
electromagnetic field-strength tensor and denote the time-like
components~$C^{0i}$ by~$\vec{E}$ and the space-like
components~$C^{ij}$ by~$\vec{B}$.  Instead of constructing NCQFT
directly in terms of the operators $\hat{x}$, we encode the NC
structure~(\ref{eq:NC-space-time}) of space-time by means of a
deformed product of functions on an ordinary commuting space-time, the
so called Moyal-Weyl $\star$-product:
\begin{equation}\label{eq:Moyal-Weyl}
  f(x)\star g(x) = f(x) \ee^{ \frac{\ii}{2}\overleftarrow{\partial^\mu}
      \theta_{\mu\nu}
      \overrightarrow{\partial^\nu}} g(x)\,.
\end{equation}

For the implementation of the gauge structure of the SM, we use the
framework introduced in~\cite{NCSM}, where the Lie algebra valued
gauge and matter fields~$A_\xi$ and~$\psi$ are mapped to universal
enveloping algebra valued fields~$\hat{A}_\xi[A,\theta]$
and~$\hat{\psi}[A,\psi,\theta]$, allowing the $SU(N)$~gauge groups and
fractional $U(1)$-charges of the SM on NC space-time.  These so-called
Seiberg Witten Maps~(SWM) are defined as solutions of the following gauge
equivalence equations, that express the requirement that the NC gauge
transformations are realized by ordinary gauge transformations:
\begin{subequations}
\label{eq:SWM-condition-infinitesimal}
\begin{align}
\label{eq:SWM-condition-infinitesimal-A}
  \hat\delta_\alpha \hat A_\mu(A,\theta)
    &= \delta_\alpha\hat A_\mu(A,\theta) \\
\label{eq:SWM-condition-infinitesimal-psi}
  \hat\delta_\alpha \hat \psi(\psi,A,\theta)
    &= \delta_\alpha\hat\psi(\psi,A,\theta)\,.
\end{align}
\end{subequations}
The solutions of~(\ref{eq:SWM-condition-infinitesimal}) can be
obtained as an expansion in powers of~$\theta$.  While we have
constructed the most general second order expressions
recently~\cite{Alboteanu:2007bp}, we will restrict ourselves here to
the first order in~$\theta$ to be consistent with the existing LHC
study~\cite{Alboteanu/Ohl/Rueckl:2006}.

The construction sketched in the previous paragraph introduces
momentum dependent corrections to the SM vertices, as well as new
vertices that are absent in the SM, e.\,g.{} $f\bar{f}VV$~contact
interactions among fermions and gauge bosons.  In addition, the gauge
boson sector of the NCSM shows a new feature, characteristic to the
universal algebra valued approach~\cite{NCSM}. The action depends on
the choice of the representation, resulting in different versions of
the model: the minimal NCSM containing no triple couplings among
neutral gauge bosons and the nonminimal NCSM, where such triple gauge
boson~(TGB) couplings, that are
forbidden in the SM, appear. The coupling strength of TGB interactions
are not uniquely fixed in the nonminimal NCSM, but constrained to a
finite domain (see Figure~\ref{Fig:1}, left).  An important aspect of
our phenomenological analysis is probing different values of these
couplings at the ILC and deriving the corresponding sensitivity on the
NC scale~$\Lnc$.  This will reveal a complementarity with measurements
at the~LHC.

\section{Phenomenology}
We perform a phenomenological analysis of the unpolarized scattering
process~$e^+e^-\to Z \gamma$ in the minimal as well as in the
nonminimal NCSM. The final state was selected to contain a $Z$-boson,
since the axial coupling of the~$Z$ is crucial for a non-cancellation
of the NC effects after summing over
polarizations~\cite{Ohl/Reuter:2004:NCPC,Alboteanu/Ohl/Rueckl:2006}.

In the minimal NCSM, the $\mathcal{O}(\theta)$ contribution to the
$e^+e^-\to Z\gamma$ scattering amplitude is given by the diagrams
\begin{center}
  \includegraphics{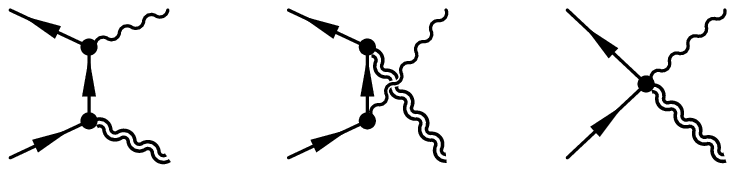}\,,
\end{center}
whereas in the nonminimal NCSM two additional $s$-channel diagrams
\begin{center}
  \includegraphics{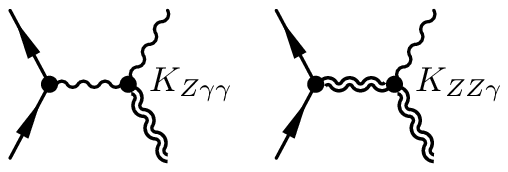}
\end{center}
have to be added, introducing a dependence on~$K_{Z\gamma\gamma}$
and~$K_{ZZ\gamma}$.

\subsection{Dependence on the Azimuthal Angle}
\label{sec:Azimuthal-Dependence}
A NC structure of space-time as introduced
in~(\ref{eq:NC-space-time}), breaks Lorentz invariance, including
rotational invariance around the beam axis. This leads to a dependence
of the cross section on the azimuthal angle, that is otherwise absent
in the SM, as well as in most other models of physics beyond the
SM (see Figure~\ref{Fig:1}, right).  In principle, we can distinguish
$\vec{E}$-type and $\vec{B}$-type NC contributions by their different
dependence on the polar scattering angle: the differential cross
section is antisymmetric in~$\cos\vartheta$ for~$\vec{E}\neq 0$ and it
is symmetric for $\vec{B}\neq 0$.  However, the dependence of the
cross section on~$\vec{E}$ is much stronger than the one on~$\vec{B}$,
which will make it very hard to discover the latter at the
LHC~\cite{Alboteanu/Ohl/Rueckl:2006}.

\subsection{Dependence on the Coupling Constants}
\label{sec:CC-Dependence}
\begin{figure}
  \begin{center}
    \includegraphics[width=.45\textwidth]{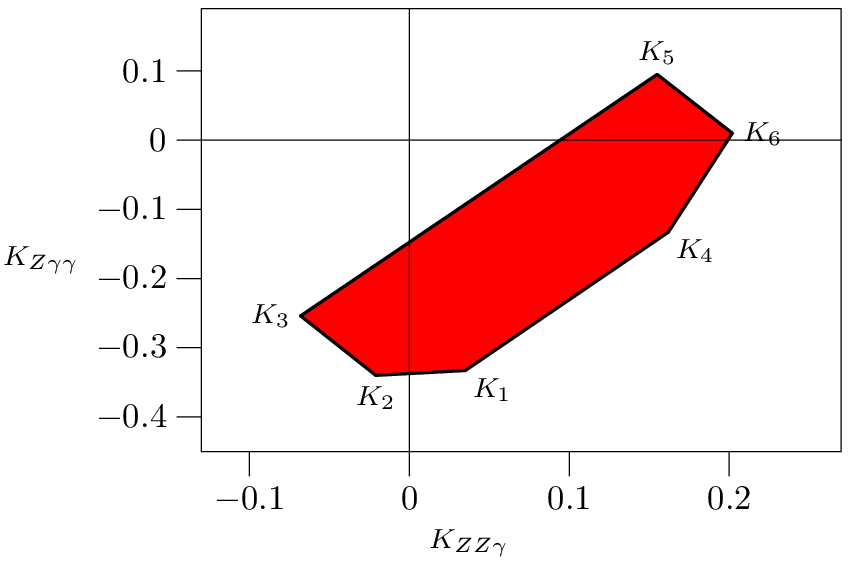}\quad
    \includegraphics[width=.45\textwidth]{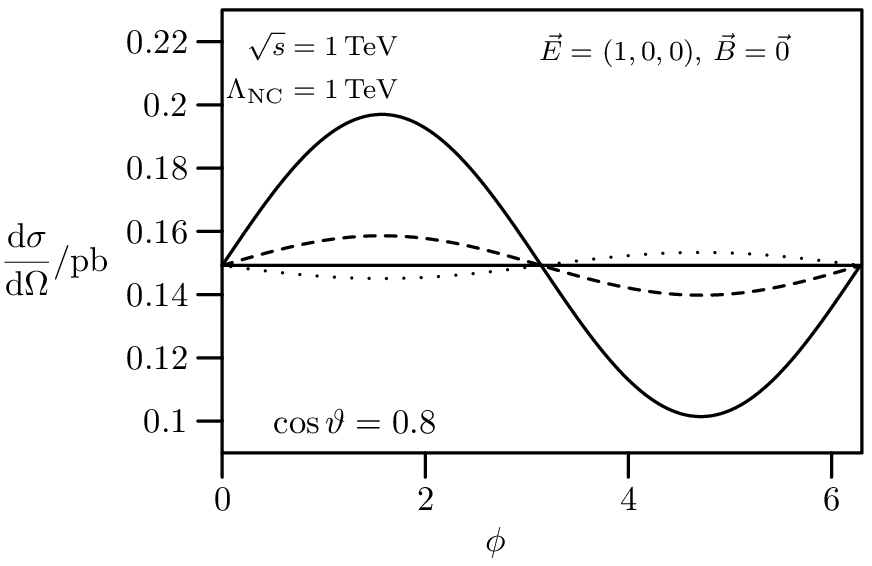}
  \end{center}
\caption{Left: The allowed region for the values of the couplings
  $K_{Z\gamma\gamma}$ and $K_{ZZ\gamma}$ in the nonminimal NCSM.  
  Right: Azimuthal dependence
  of the cross section for~$e^+e^-\to Z \gamma$, 
  in the nonminimal NCSN with different values for the TGB couplings:
  $K_1 = (-0.333,0.035)$ (solid) and $K_5 = (0.095,0.155)$ (dotted), 
  and in the minimal NCSM (dashed).}
\label{Fig:1}
\end{figure}

Since the $t$- and $u$-channel diagrams as well as the contact term 
are proportional to $Q^2$, $Q$ being the particle charge in the 
initial state, the cross section in the minimal NCSM depends only 
on the modulus~$|Q|$.
In contrast, in the nonminimal NCSM, the interference
with the $s$-channel diagrams adds a $Q^3$~term to the cross section
and the cross section also depends on $\text{sgn}(Q)$.  As a result,
NC effects in $e^+e^-\to Z\gamma$ are maximally enhanced by the
$s$-channel contribution for the pairs of couplings $K_1$ and $K_2$,
corresponding to the lower edge of the polygon in Figure~\ref{Fig:1},
left. However, the same couplings lead to cancellations of the NC
effects for $u\bar{u}$ scattering resulting in minimal deviations of
the NCSM with respect to the SM. In this sense, the ILC will nicely
complement the LHC.  On the other hand, the pair of couplings $K_5$,
which produces maximal effects at the LHC, will lead to an NCSM cross
section comparable to the one where the TGB couplings vanish.

\subsection{Monte Carlo Simulations for the ILC}
In order to estimate the sensitivity of the ILC on the NC
scale $\Lnc$, we have performed Monte Carlo simulations using the event
generator WHIZARD~\cite{Kilian:2007gr}. In the analysis we used a
center of mass energy of $\sqrt{s}=\unit[500]{GeV}$ and an integrated
luminosity of $\mathcal{L} = \unit[500]{fb^{-1}}$.

\begin{figure}
  \begin{center}
    \includegraphics[width=.45\textwidth]{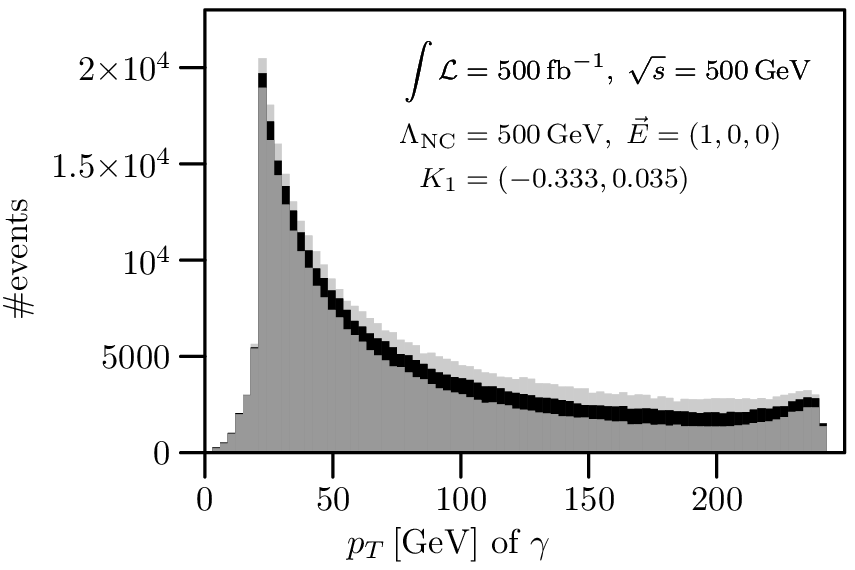}\quad
    \includegraphics[width=.45\textwidth]{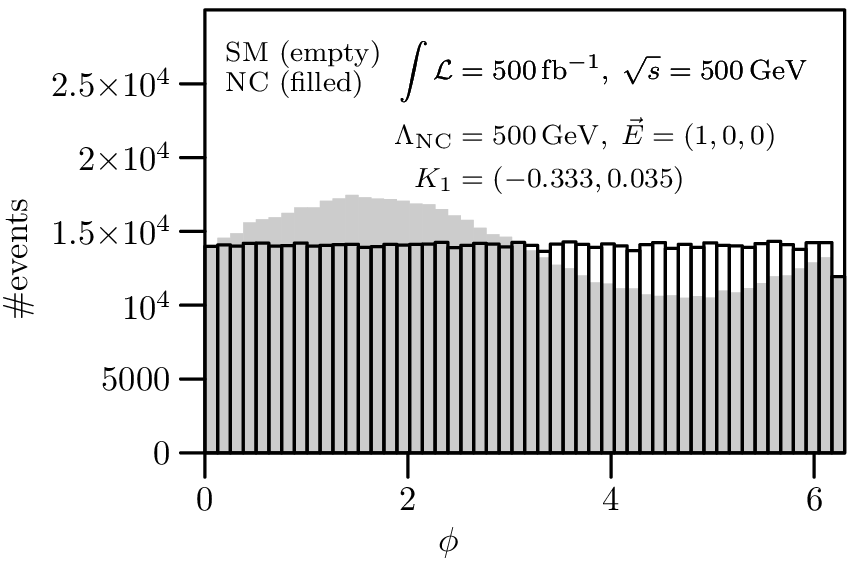}
  \end{center}
\caption{Left: Monte Carlo simulation for the photon $p_T$ distribution 
         in the process $ e^+e^-\to Z \gamma$ at the ILC showing 
         the distribution in the NCSM for $0.0<\phi<\pi$ ($\pi<\phi<2\pi$)
         above (below) the black SM histogramm. 
  Right: Monte Carlo simulation for the azimuthal dependence of the
         process $ e^+e^-\to Z \gamma$ at the ILC.}   
\label{Fig:2}
\end{figure}

A typical signature for new physics is a modified
$p_T$-distribution. Previously, we have studied $pp\to Z\gamma\to
e^+e^-\gamma$ at the LHC and the deviation from the SM $p_T(\gamma)$
distribution could not be resolved due to the poor statistics and
complicated cuts~\cite{Alboteanu/Ohl/Rueckl:2006}.  However, the high
statistics and the clean initial state of the ILC, allows deviations
of the NCSM from the SM to be seen also in the $p_T$ distribution for
reasonable values of $\Lnc$ (see Figure~\ref{Fig:2}, left).  Of
course, cuts with respect to the azimuthal angle~$\phi$ have to be
applied, because otherwise all $\mathcal{O}(\theta)$ interference
effects will cancel, since the events ``missing'' in one hemisphere
(e.\,g.~for $\pi<\phi<2\pi$) are compensated by the ``excess'' of
events in the other. Figure~\ref{Fig:2}, right, shows this
distribution exemplarily, where for the TGB couplings we have chosen
the set of values, for which we expect the largest deviation from the
SM distribution in electron-positron scattering,
i.\,e.~$K_1$.

As shown in~\cite{Alboteanu/Ohl/Rueckl:2006}, the strong boost along
the beam axis from the partonic to the hadronic CMS at the LHC induces
kinematical correlations between $(E_1,B_2)$ and $(E_2,B_1)$,
respectively. Thus, in the laboratory frame we always deal with an
entanglement of time- and space-like noncommutativity. Fortunately,
the different properties of the $\vec{E}$ and $\vec{B}$ parameters
with respect to the partonic scattering angle discussed in
section~\ref{sec:Azimuthal-Dependence} allows separate measurements of
the time- and space-like components of~$\theta$. Integrating just over
one hemisphere (i.\,e.~$-0.9<\cos\vartheta^*<0$
or~$0<\cos\vartheta^*<0.9$) we can perform a measurement of~$\vec{E}$,
since the $\vec{B}$~dependence is negligibly small. On the other hand,
an integration over the whole sphere
(i.\,e.~$-0.9<\cos\vartheta^*<0.9$) in principle provides a pure
measurement of $\vec{B}$, since the effect of $\vec{E}$ will
completely cancel out, due to its antisymmetry.

One advantage of the ILC compared to the LHC is the only mildly
boosted initial state. We have an $e^+e^-$ initial state, where only 
beamstrahlung has to be accounted for, which we have done, using
CIRCE~\cite{Ohl:1996fi} inside WHIZARD~\cite{Kilian:2007gr}. This will
lead to a boost of the CMS of the electrons to the laboratory
frame. Yet, compared to the LHC, this boost is negligibly small:
$\langle |\beta_{\text{ILC}}|\rangle = 0.14$ versus 
$\langle |\beta_{\text{LHC}}|\rangle = 0.8$.  We
therefore have negligible correlations between $E_1$ and~$B_2$ or
$E_2$ and~$B_1$, respectively, and we can derive the bounds on $\Lnc$
separately for the case of purely $\vec{E}$ or purely $\vec{B}$
noncommutativity.

\section{Results and Conclusions}
Focussing on the azimuthal dependency (Figure
\ref{Fig:2}) we have performed likelihood fits similar to the ones 
described in~\cite{Alboteanu/Ohl/Rueckl:2006} in order to derive bounds on the
NC scale~$\Lnc$. The results are summarized in Table~\ref{Tab:ILC-bounds}.
\begin{table}
\renewcommand{\arraystretch}{1.5}
\begin{center}
  \begin{tabular}{|c||c|c|}\hline 
    $(K_{Z\gamma\gamma},K_{ZZ\gamma})$ & $|\vec{E}|^2 = 1, \vec{B}=0$   & $\vec{E}=0, |\vec{B}|^2 = 1$ \\    \hline 
    $K_0\equiv (0,0)$          & $\Lnc \gtrsim \unit[2 ]{TeV}$  & $\Lnc \gtrsim \unit[0.4]{TeV}$ \\ \hline
    $K_1\equiv (-0.333,0.035)$  & $\Lnc \gtrsim \unit[5.9]{TeV}$ & $\Lnc \gtrsim \unit[0.9]{TeV}$ \\ \hline 
    $K_5\equiv (0.095,0.155)$   & $\Lnc \gtrsim \unit[2.6]{TeV}$   & $\Lnc \gtrsim \unit[0.25]{TeV}$ \\ \hline 
    $K_3\equiv (-0.254,-0.048)$ & $\Lnc \gtrsim \unit[5.4]{TeV}$   & $\Lnc \gtrsim \unit[0.9]{TeV}$ \\ \hline 
  \end{tabular}
\end{center}
\renewcommand{\arraystretch}{1}
\caption{
  Bounds on $\Lnc$ from $pp\to Z\gamma\to e^+e^-\gamma$ at the LHC, for the minimal (first row) and nonminimal NCSM}
\label{Tab:ILC-bounds}
\end{table}
In contrast to the LHC case, the ILC is sensitive on all
noncommutative parameters, time-like and space-like, as well as on all
values of the TGB couplings. The ILC is especially sensitive on the
couplings lying in the lower region of the polygon of Figure
\ref{Fig:1}. These are exactly the set of TGB couplings for which the
LHC is less sensitive, while the TGB couplings leading to maximal
deviations at the LHC, lead to minimal effects at the ILC.  Thus,
probing the TGB couplings at the ILC is complementary to searches at
the LHC.  If a noncommutative structure of space-time exists in nature
at a scale of the order of 1 TeV without being discovered at the LHC
because of an unfavorable value of the TGB coupling (i.\,e.~in the
upper part of the polygon in Figure~\ref{Fig:1}), then the ILC will
see it.

\section{Acknowledgments}
This research is supported by Deutsche Forschungsgemeinschaft (grant
RU 311/1-1 and Research Training Group 1147 \textit{Theoretical
Astrophysics and Particle Physics}) and by Bundesministerium f\"ur
Bildung und Forschung BMBF (grant 05H4\-WWA/2).  A.\,A.~gratefully
acknowledges support from Evangelisches Studienwerk e.\,V.~Villigst.
A.\,A.~thanks the members of SLAC Theory Group for their kind
hospitality.


\begin{footnotesize}

\end{footnotesize}



\begin{thebibliography}{99}
\bibitem{url} Slides: \\ 
\verb$http://ilcagenda.linearcollider.org/contributionDisplay.py?contribId=237&sessionId=72&confId=1296$

\bibitem{Alboteanu/Ohl/Rueckl:2006}
  A.~Alboteanu, T.~Ohl and R.~R\"uckl,
  Phys.{} Rev.{} \textbf{D74}, 096004 (2006)
  [arXiv:hep-ph/0608155];
  PoS \textbf{HEP2005} (2006), 322
  [arXiv:hep-ph/0511188].

\bibitem{NCSM}
  X.~Calmet, B.~Jur{\v c}o, P.~Schupp, J.~Wess and M.~Wohlgenannt,
  Eur.{} Phys.{} J.{} \textbf{C23}, 363 (2002) 
  [arXiv:hep-ph/0111115];
  B.~Melic, K.~Passek-Kumericki, J.~Trampetic, P.~Schupp and M.~Wohlgenannt,
  Eur.{} Phys.{} J.{}  C \textbf{42}, 483 (2005)
  [arXiv:hep-ph/0502249];
  B.~Melic, K.~Passek-Kumericki, J.~Trampetic, P.~Schupp and M.~Wohlgenannt,
  Eur.{} Phys.{} J.{}  C \textbf{42}, 499 (2005)
  [arXiv:hep-ph/0503064].

\bibitem{Alboteanu:2007bp}
  A.~Alboteanu, T.~Ohl and R.~R\"uckl,
  arXiv:0707.3595 [hep-ph].

\bibitem{Ohl/Reuter:2004:NCPC}
  T.~Ohl and J.~Reuter,
  Phys.{} Rev.{} \textbf{D70}, 076007 (2004)
  [arXiv:hep-ph/0406098];
  T.~Ohl and J.~Reuter
  [arXiv:hep-ph/0407337].

\bibitem{Kilian:2007gr}
  W.~Kilian, T.~Ohl and J.~Reuter,
  arXiv:0708.4233 [hep-ph];
  M.~Moretti, T.~Ohl and J.~Reuter,
  [arXiv:hep-ph/0102195];
  W.~Kilian,
  LC-TOOL-2001-039.

\bibitem{Ohl:1996fi}
  T.~Ohl, 
  Comput.{} Phys.{} Commun.{} \textbf{101}, 269 (1997)
  [arXiv:hep-ph/9607454].


\end{thebibliography}
\end{document}